\documentclass[prd,superscriptaddress,amsfonts,amssymb,amsmath,showpacs,showkeys]{revtex4}

\usepackage{graphics}
\usepackage{eurosym}
\usepackage{amsfonts}
\usepackage{amssymb}
\usepackage{amsmath}
\usepackage{graphicx}
\usepackage[font={footnotesize,it}]{caption}
\usepackage{hyperref}

\def\m{\mu}
\def\n{\nu}

\begin{document}

\title{Tunneling of Massive/Massless Bosons From the Apparent Horizon of FRW Universe}
\author{Kimet Jusufi}
\email{kimet.jusufi@unite.edu.mk}
\affiliation{Institute of Physics, Faculty of Natural Sciences and Mathematics, Ss. Cyril and Methodius University, Arhimedova 3, 1000 Skopje, Macedonia}
\affiliation{Physics Department, State University of Tetovo, Ilinden Street nn, 1200, Tetovo,
Macedonia}

\author{Ali Ovgun}
\email{ali.ovgun@emu.edu.tr}
\affiliation{Instituto de F\'{\i}sica, Pontificia Universidad Cat\'olica de
Valpara\'{\i}so, Casilla 4950, Valpara\'{\i}so, Chile}
\affiliation{Physics Department, Eastern Mediterranean University, Famagusta,
Northern Cyprus, Turkey}

\author{Gordana Apostolovska}
\email{gordanaa@pmf.ukim.mk}
\affiliation{Institute of Physics, Faculty of Natural Sciences and Mathematics, Ss. Cyril and Methodius University, Arhimedova 3, 1000 Skopje, Macedonia}

\date{\today }

\begin{abstract}
In this article we investigate the Hawking radiation of vector particles from the apparent horizon of a Friedmann-Robertson-Walker (FRW) universe in the framework of quantum tunnelling method. Furthermore we use Proca equation, a relativistic wave equation for a massive/massless spin-1 particle (massless $\gamma$ photons, weak massive $W^{\pm}$ and $Z^{0}$ bosons, strong massless gluons and $\rho$ and $\omega$ mesons) together with a Painlev\'{e} spacetime metric for the FRW universe. We solve the Proca equation via Hamilton-Jacobi (HJ) equation and the WKB approximation method. We recover the same result for the Hawking temperature associated with vector particles as in the case of scalar and Dirac particles tunnelled from outside to the inside of the apparent horizon in a FRW universe. 
\end{abstract}
\maketitle

\section{Introduction}

In early 1970s, Stephen Hawking came with an idea that black holes emit thermal radiation because of the quantum effects near the event horizon \cite{hawking1,hawking2,hawking3}. It means that black holes are not completely black. This may meet a interesting experimental base for searching the vertex of general relativity and quantum mechanics. Unfortunately, the temperatures of Hawking radiation from black holes
are far too small to be observed within
the near future. 

Since Hawking, many derivations of this radiation have been made. Most famous derivation is made by Parikh and Wilczek in which they use a tunnelling process with a semi classical WKB approximation \cite{perkih1,perkih2,perkih3,kraus1,kraus2,ang,sri1,sri2,vanzo}.  A number of important contributions have been done using different types of spherically symmetric and stationary black holes, for different types fields and corresponding particles such as  scalar as well as for Dirac particles \cite{mann0,mann1,mann2,mann3,xiang1,kruglov1,sakalli1,sakalli2,a1,a2,a3,k1,k2,k3}.

Besides the progress in the field, there are a number of conceptual problems related to Hawking radiation. For example, the information loss paradox remains still unsolved. Furthermore there is no agreement among physicist whether this radiation is produced near the event horizon or, in some larger region surrounding the black hole (see, for example \cite{giddings}).  If we assume that Hawking quanta are located very close to the event horizon, there is a problem when considering the infinite blueshift. In fact, it's interesting to note that according to the tunnelling method this process took place near the event horizon.  However, recent studies show that quantum gravity effects and the nature of particles may affect the Hawking temperature. That is to say, when the GUP effects are introduced the Hawking quanta near the event horizon can never be infinite-blue shifted due to the existence of minimal length \cite{xiang2}.

Recently the quantum tunnelling of scalar and Dirac particles from the apparent horizon in a FRW universe has been investigated in Refs. \cite{frw1,frw2,frw3,frw4}. On the other hand, since the famous thermodynamical derivation of Einstein's field equation by Jacobson \cite{jacobson}, thermodynamical  aspects of the apparent horizon in  for FRW universe has been widely studied in the context of  different gravity theories \cite{frw7}. In this paper, we shall extend the work of Cai {\it{et al}} \cite{frw3} to explore the Hawking radiation of massive/massless spin-1 particles in the framework of quantum tunnelling method to compute Hawking temperature from the apparent horizon of a FRW universe.

This paper consists of three main parts. In Section II, we give a brief review and introduce a Painlev\'{e}-type spacetime metric for the FRW universe and then investigate the Hawking radiation of massive bosons via tunnelling. In Section III, we apply the same method, namely the HJ equation to the massless Hawking quanta (photon particles) and derive the Hawking temperature from the apparent horizon of a FRW universe. Finally, in the last part, we conclude our paper.

\section{Hawking Radiation of massive bosons via Tunnelling}

The FRW Universe can be described by the following metric
\begin{equation}
ds^2=-dt^2+a^{2}(t)\left(\frac{dr^2}{1-k r^2}+r^2 d\theta^2 +r^2 \sin^2{\theta} d\varphi^2 \right).
\end{equation}
in which $a(t)$ is the scale factor, and $k$ encodes the geometry of FRW universe. In particular,  $k=-1,0,1$, corresponds to a closed, flat, and open universe, respectively. Introducing a new coordinate transformation $\tilde{r}=r a$, the above metric reduces to
\begin{equation}
ds^2=h_{ab}dx^a dx^b+\tilde{r}^2 (d\theta^2+\sin^2 \theta d\varphi^2).
\end{equation}

From this equation we can define the apparent horizon as follows
\begin{equation}
h^{ab}\partial_{a} \tilde{r}\partial_{b}\tilde{r}=0.
\end{equation}

Then one can show the following relation for the apparent horizon 
\begin{equation}
\tilde{r}_{A}=\frac{1}{\sqrt{H^2+k/a^2}},
\end{equation}
where $H=\dot{a}/a$. Using $\tilde{r}=r a$, one can find a Painlev\'{e}-type coordinates for FRW universe metric as follows
\begin{equation}
ds^2=-\frac{1-\tilde{r}^2/{\tilde{r}_{A}}^2}{1-k\tilde{r}^2/a^2}dt^2-\frac{2H\tilde{r}}{1-k\tilde{r}^2/a^2}dt  d\tilde{r}+\frac{1}{1-k\tilde{r}^2/a^2}d\tilde{r}^2+\tilde{r}^2 (d\theta^2+\sin^2 \theta d\varphi^2).\label{5}
\end{equation}

Later on, we shall use this metric to study the tunnelling of massive/massless bosons. Particles are nothing else but excitations of some fields, in particular vector particles are excitations of vector fields $\Psi^{\mu}$ which are described by the relativistic wave equation known as Proca equation
\begin{equation}
\frac{1}{\sqrt{-g}}\partial_{\mu}\left(\sqrt{-g}\Psi^{\nu\mu}\right)+\frac{m^{2}}{\hbar^2}\Psi^{\nu}=0,\label{6}
\end{equation}

in order to solve the Proca equation we shall employ the following ansatz for our vector field $\Psi^{\nu}$, as follows
\begin{equation}
\Psi_{\mu}=C_{\mu} \exp \left(\frac{i}{\hbar} S(t,\tilde{r},\theta, \varphi) \right).\label{7}
\end{equation}

On the other hand an observer will detect Hawking quanta. Before we shall go on to separate the variables in the action for the Hawking quanta one must consider the Kodama vector. The physical significance of the Kodama vector is related to the energy of those Hawking quanta measured by the Kodama observer
\begin{equation}
\omega=-K^a \partial_{a}S=-\sqrt{1-k \tilde{r}^2/a^2}\,\partial_{t}S,
\end{equation}
in which the Kodama vector is defined via
\begin{equation}
K^a=\sqrt{1-k \tilde{r}^2/a^2} \,\left(\frac{\partial}{\partial t}\right)^a.
\end{equation}

In terms of the above relations one may choose the following form to the action for the Hawking quanta
\begin{equation}
S\left(t,\tilde{r},\theta, \varphi \right)=-\int \frac{\omega }{\sqrt{1-k \tilde{r}^2/a^2}} dt+R(\tilde{r})+P(\theta, \varphi).\label{10}
\end{equation}

Substituting our action \eqref{10}, into the Eqs. \eqref{6}, \eqref{7} in a Painlev\'{e}-type spacetime metric \eqref{5} for the FRW universe, we find the following set of four differential equations:
\bigskip
\begin{eqnarray}\notag
0&=&-\left[\frac{g(\tilde{r})\sin^{2}\theta (\partial_{\theta}P)^2+g(\tilde{r}) (\partial_{\varphi}P)^2+\sin^{2}\theta \tilde{r}^2\Big( g(\tilde{r}) m^2-t E(\tilde{r})E(\tilde{r})^{\prime}+E(\tilde{r})R(\tilde{r})^{\prime}\Big)}{\Delta(\tilde{r}) \sin^{2}\theta \tilde{r}^2}\right]C_1\\\notag
&-& \left[\frac{g(\tilde{r})E(\tilde{r})^{\prime} t-g(\tilde{r})R(\tilde{r})^{\prime}+ h(\tilde{r})E(\tilde{r})}{\Delta(\tilde{r}) \tilde{r}^2}\right](\partial_{\theta}P) C_2-\left[\frac{g(\tilde{r})E(\tilde{r})^{\prime} t-g(\tilde{r})R(\tilde{r})^{\prime}+ h(\tilde{r})E(\tilde{r})}{\Delta(\tilde{r})\sin^{2}\theta \tilde{r}^2}\right](\partial_{\varphi}P) C_3 \\
&-& \left[\frac{h(\tilde{r})\sin^{2}\theta (\partial_{\theta}P)^2+h(\tilde{r}) (\partial_{\varphi}P)^2+\sin^{2}\theta \tilde{r}^2\Big( t^2 {E(\tilde{r})^{{\prime}}}^2 +h(\tilde{r})m^2-2 t R(\tilde{r})^{\prime}E(\tilde{r})^{\prime}+{R(\tilde{r})^{\prime}}^2\Big)}{\Delta(\tilde{r}) \sin^{2}\theta \tilde{r}^2}\right]C_4
\end{eqnarray}

\begin{eqnarray}\notag
0&=& \left[\frac{f(\tilde{r})\sin^{2}\theta (\partial_{\theta}P)^2+f(\tilde{r}) (\partial_{\varphi}P)^2+\sin^2 \theta \tilde{r}^2 \Big( f(\tilde{r}) m^2-E(\tilde{r})^2\Big)}{\Delta(\tilde{r}) \sin^{2}\theta \tilde{r}^2}\right]C_1\\\notag
&+&  \left[\frac{f(\tilde{r})E(\tilde{r})^{\prime} t-f(\tilde{r})R(\tilde{r})^{\prime}-g(\tilde{r})E(\tilde{r})}{\Delta(\tilde{r}) \tilde{r}^2}\right](\partial_{\theta}P) C_2+\left[\frac{f(\tilde{r})E(\tilde{r})^{\prime} t-f(\tilde{r})R(\tilde{r})^{\prime}-g(\tilde{r})E(\tilde{r})}{\Delta(\tilde{r}) \sin^2 \theta \tilde{r}^2}\right](\partial_{\varphi}P) C_3\\
&-&\left[\frac{g(\tilde{r})\sin^{2}\theta (\partial_{\theta}P)^2+g(\tilde{r}) (\partial_{\varphi}P)^2+\sin^{2}\theta \tilde{r}^2\Big( g(\tilde{r}) m^2-t E(\tilde{r})E(\tilde{r})^{\prime}+E(\tilde{r})R(\tilde{r})^{\prime}\Big)}{\Delta(\tilde{r}) \sin^{2}\theta \tilde{r}^2}\right]C_4,
\end{eqnarray}

\begin{eqnarray}\notag
0&=& \left[\frac{f(\tilde{r})E(\tilde{r})^{\prime} t-f(\tilde{r})R(\tilde{r})^{\prime}-g(\tilde{r})E(\tilde{r})}{\Delta(\tilde{r}) \tilde{r}^2}\right](\partial_{\theta}P) C_1\\\notag
&+& \left[\frac{\Delta (\partial_{\theta}P)^2+\sin^{2}\theta \tilde{r}\Big(t^2 f (E(\tilde{r})^{\prime})^2-2t E(\tilde{r})^{\prime} (gE+fR(\tilde{r})^{\prime})+m^2 \Delta +2g E R(\tilde{r})^{\prime}-E^2 h+f (R(\tilde{r})^{\prime})^2   \Big) }{\Delta(\tilde{r}) \sin^{2}\theta \tilde{r}^4}\right]C_2\\
&-& \frac{(\partial_{\theta}P (\partial_{\varphi}P)}{r^4 \sin^2 \theta}C_3+\left[\frac{g(\tilde{r})E(\tilde{r})^{\prime} t-g(\tilde{r})R(\tilde{r})^{\prime}+h(\tilde{r})E(\tilde{r})}{\Delta(\tilde{r}) \tilde{r}^2}\right](\partial_{\theta}P) C_4,
\end{eqnarray}

\begin{eqnarray}\notag
0&=&\left[\frac{f(\tilde{r})E(\tilde{r})^{\prime} t-f(\tilde{r})R(\tilde{r})^{\prime}-g(\tilde{r})E(\tilde{r})}{\Delta(\tilde{r})\sin^{2}\theta \tilde{r}^2}\right](\partial_{\varphi}P) C_1-\frac{(\partial_{\theta}P (\partial_{\varphi}P)}{r^4 \sin^2 \theta}C_2\\\notag
&+& \left[\frac{\Delta (\partial_{\theta}P)^2+ \tilde{r}\Big(t^2 f (E(\tilde{r})^{\prime})^2-2t E(\tilde{r})^{\prime} (gE+fR(\tilde{r})^{\prime})+m^2 \Delta +2g E R(\tilde{r})^{\prime}-E^2 h+f (R(\tilde{r})^{\prime})^2   \Big) }{\Delta(\tilde{r}) \sin^{2}\theta \tilde{r}^4}\right]C_3\\
&-&\left[\frac{g(\tilde{r})E(\tilde{r})^{\prime} t-g(\tilde{r})R(\tilde{r})^{\prime}+ h(\tilde{r})E(\tilde{r})}{\Delta(\tilde{r})\sin^{2}\theta \tilde{r}^2}\right](\partial_{\varphi}P) C_4,
\end{eqnarray}
where we have introduced 
\begin{eqnarray}
f(\tilde{r})&=&\frac{1-\tilde{r}^2/{\tilde{r}_{A}}^2}{1-k\tilde{r}^2/a^2},\\
g(\tilde{r})&=&\frac{H\tilde{r}}{1-k\tilde{r}^2/a^2},\\
h(\tilde{r})&=&\frac{1}{1-k\tilde{r}^2/a^2},\\
E(\tilde{r})&=&\frac{\omega}{\sqrt{1-k\tilde{r}^2/a^2}},\\
\Delta(\tilde{r})&=&f(\tilde{r}) h(\tilde{r})+g(\tilde{r})^2.
\end{eqnarray}

From those four equations it follows a matrix equation using a $4\times 4$ matrix $\Xi$ and multiplied by the transpose of a vector $(C_{1}, C_{2}, C_{3}, C_{4})$. This  matrix equation reads
\begin{equation}
\Xi(C_{1}, C_{2}, C_{3}, C_{4})^{T}=0.
\end{equation}

Moreover this matrix is characterised by the following non--zero matrix elements:

\begin{eqnarray}\nonumber
\Xi_{11}&=&\Xi_{24}=-\frac{g(\tilde{r})\sin^{2}\theta (\partial_{\theta}P)^2+g(\tilde{r}) (\partial_{\varphi}P)^2+\sin^{2}\theta \tilde{r}^2\Big( g(\tilde{r}) m^2-t E(\tilde{r})E(\tilde{r})^{\prime}+E(\tilde{r})R(\tilde{r})^{\prime}\Big)}{\Delta(\tilde{r}) \sin^{2}\theta \tilde{r}^2},\\\nonumber
\Xi_{12}&=&-\Xi_{34}=-\left[\frac{g(\tilde{r})E(\tilde{r})^{\prime} t-g(\tilde{r})R(\tilde{r})^{\prime}+ h(\tilde{r})E(\tilde{r})}{\Delta(\tilde{r}) \tilde{r}^2}\right](\partial_{\theta}P),\\\nonumber
\Xi_{13}&=&\Xi_{44}=-\left[\frac{g(\tilde{r})E(\tilde{r})^{\prime} t-g(\tilde{r})R(\tilde{r})^{\prime}+ h(\tilde{r})E(\tilde{r})}{\Delta(\tilde{r})\sin^{2}\theta \tilde{r}^2}\right](\partial_{\varphi}P),\\\nonumber
\Xi_{14}&=&-\frac{h(\tilde{r})\sin^{2}\theta (\partial_{\theta}P)^2+h(\tilde{r}) (\partial_{\varphi}P)^2+\sin^{2}\theta \tilde{r}^2\Big( t^2 {E(\tilde{r})^{{\prime}}}^2 +h(\tilde{r})m^2-2 t R(\tilde{r})^{\prime}E(\tilde{r})^{\prime}+{R(\tilde{r})^{\prime}}^2\Big)}{\Delta(\tilde{r}) \sin^{2}\theta \tilde{r}^2},\\\nonumber
\Xi_{21}&=&\frac{f(\tilde{r})\sin^{2}\theta (\partial_{\theta}P)^2+f(\tilde{r}) (\partial_{\varphi}P)^2+\sin^2 \theta \tilde{r}^2 \Big( f(\tilde{r}) m^2-E(\tilde{r})^2\Big)}{\Delta(\tilde{r}) \sin^{2}\theta \tilde{r}^2} ,\\\nonumber
\Xi_{22}&=& \Xi_{31}=\left[\frac{f(\tilde{r})E(\tilde{r})^{\prime} t-f(\tilde{r})R(\tilde{r})^{\prime}-g(\tilde{r})E(\tilde{r})}{\Delta(\tilde{r}) \tilde{r}^2}\right](\partial_{\theta}P),\\\nonumber
\Xi_{23}&=&\Xi_{41}=\left[\frac{f(\tilde{r})E(\tilde{r})^{\prime} t-f(\tilde{r})R(\tilde{r})^{\prime}-g(\tilde{r})E(\tilde{r})}{\Delta(\tilde{r}) \sin^2 \theta \tilde{r}^2}\right](\partial_{\varphi}P),\\\nonumber
\Xi_{32}&=& \frac{\Delta (\partial_{\theta}P)^2+\sin^{2}\theta \tilde{r}\Big(t^2 f (E(\tilde{r})^{\prime})^2-2t E(\tilde{r})^{\prime} (gE+fR(\tilde{r})^{\prime})+m^2 \Delta +2g E R(\tilde{r})^{\prime}-E^2 h+f (R(\tilde{r})^{\prime})^2   \Big) }{\Delta(\tilde{r}) \sin^{2}\theta \tilde{r}^4},\\\nonumber
\Xi_{33}&=&\Xi_{42}=-\frac{(\partial_{\theta}P (\partial_{\varphi}P)}{r^4 \sin^2 \theta}\,\\\nonumber
\Xi_{43}&=& \frac{\Delta (\partial_{\theta}P)^2+ \tilde{r}\Big(t^2 f (E(\tilde{r})^{\prime})^2-2t E(\tilde{r})^{\prime} (gE+fR(\tilde{r})^{\prime})+m^2 \Delta +2g E R(\tilde{r})^{\prime}-E^2 h+f (R(\tilde{r})^{\prime})^2   \Big) }{\Delta(\tilde{r}) \sin^{2}\theta \tilde{r}^4}.
\end{eqnarray}

In order to see the physical significance of these equations let us first solve $\det\Xi =0$, to find the following equation

\begin{eqnarray}\notag
&&\frac{1}{\sin^8 \theta \Delta(\tilde{r}) \tilde{r}^{12}}\Big[ \Big(\Big(m^2 \sin^2 \theta \tilde{r}^{2}+(\partial_{\varphi}P)^2)\Delta(\tilde{r})(\partial_{\theta}P)^2\Big)+42 m \Delta(\tilde{r}) \tilde{r}^{4} \sin^2 \theta (\partial_{\theta}P) (\partial_{\varphi}P)+ m^2 (\Delta(\tilde{r}) (\partial_{\varphi}P)^2\\
&+&\sin^2 \theta \tilde{r}^{2} \zeta(\tilde{r}))r^2\Big) \Big(\Delta(\tilde{r}) \sin^2 \theta (\partial_{\theta}P)^2 +\Delta(\tilde{r}) (\partial_{\varphi}P)^2+\sin^2 \theta \tilde{r}^{2} \zeta(\tilde{r})\Big)^2\Big]=0,\label{21}
\end{eqnarray}
where 
\begin{equation}
\zeta(\tilde{r})=\sin^2\theta \left(t^2 f(\tilde{r}) (E^{\prime})^2-2t E^{\prime} (Eg+f(\tilde{r})R^{\prime})+ m^2 \Delta(\tilde{r}) +2 g E R^{\prime}-E^2h+f(\tilde{r}) (R^{\prime})^2\right)r^2.
\end{equation}

Solving  Eq. \eqref{21} for the radial part, we find the following solution 
\begin{equation}
R(\tilde{r})_{\pm}=\frac{\omega}{\sqrt{1-k\tilde{r}^2/a^2}}\int dt - \int \frac{g(\tilde{r})E(\tilde{r})\pm\sqrt{g(\tilde{r})^2 E(\tilde{r})^2-f(\tilde{r})\Big[\Delta(\tilde{r}) \Big( m^2+(\partial_{\theta}P)^2+(\partial_{\varphi}P)^2\Big)-h(\tilde{r}) E(\tilde{r})^2\Big] }}{f(\tilde{r})}d\tilde{r},
\end{equation}
we note that $+/-$ sign in the last equation corresponds to the incoming (outgoing) wave solutions. At the apparent horizon $\tilde{r}=\tilde{r}_{A}$, it is clear that the function $f(\tilde{r})$ vanishes i.e. $f(\tilde{r}_{A})=0$. Solving this integral we see that there is a zero contribution to the imaginary part coming from the first term, on the other hand solving the second integral we find 
\begin{equation}
\text{Im}R_{+}=\pi \omega \tilde{r}_{A}, \,\,\,\,\,\text{Im}R_{-}=0.
\end{equation}

Hence, we have shown that there is no contribution of an outgoing particle
to the imaginary part of the action. This shows that the tunnelling of Hawking quanta is from the outside to the inside of the apparent horizon which is different from the black hole case.  Finally, the tunnelling probability can be calculated as \cite{frw1}
\begin{equation}
\Gamma=\frac{P_{in}}{P_{out}}=\frac{\exp\left[-2\left(\text{Im}R_{+}+\text{Im}P\right)\right]}{\exp\left[-2\left(\text{Im}R_{-}+\text{Im}P\right)\right]}=\exp\left(-2\pi \omega \tilde{r}_{A} \right).
\end{equation}

Using the last result and comparing to the Boltzmann equation $\Gamma=\exp\left(-\omega/T_{H}\right)$, we recover the Hawking temperature from the apparent horizon
\begin{equation}
T_{H}=\frac{1}{2 \pi \tilde{r}_{A} }.
\end{equation}

Hence, we have successfully derived the same relation for the Hawking temperature as in the case of scalar and Dirac particles tunnelled from the apparent horizon of FRW universe \cite{frw1,frw2,frw3,frw4}. We point out that this thermal flux of radiation with a temperature $T_{H}$ can be detected by Kodama observer which is inside the apparent horizon. 

\section{Hawking Radiation of Photons from FRW Universe}

In this section, we shall focus on tunnelling of photon particles using the HJ method \cite{samantha,Siahaan}. The action without the gauge fixing term is given by follows: 

\begin{equation}
S=-\frac{1}{4}\int F_{\m\n} F^{\m\n}\sqrt{-g}\,d^4x\label{S},
\end{equation}
and then the equation of motion for this Maxwell field is derived as 

\begin{eqnarray}
\nabla^{\m}F_{\m\n}=0,
\end{eqnarray}
where the field strength $F_{\m\n}$ is given in terms of the gauge field $A_{\m}$ as follows:
\begin{eqnarray}
F_{\m\n}&=&\nabla_{\m}A_{\n}-\nabla_{\n}A_{\m}, \label{fmunu}
\end{eqnarray}
using the Lorentz gauge condition. Now, we use the same procedure, and first we define the following ansatz for photon field (Maxwell equation)
\begin{equation}
A_{\mu}=C_{\mu} \exp \left(\frac{i}{\hbar} S(t,\tilde{r},\theta, \varphi) \right).
\end{equation}

Again the energy of those quanta (photons) measured by the Kodama observer reads
\begin{equation}
\omega=-K^a \partial_{a}S=-\sqrt{1-k \tilde{r}^2/a^2}\,\partial_{t}S,
\end{equation}
where the Kodama vector is defined via
\begin{equation}
K^a=\sqrt{1-k \tilde{r}^2/a^2} \,\left(\frac{\partial}{\partial t}\right)^a.
\end{equation}

We choose the following action for the tunnelling quanta:
\begin{equation}
S\left(t,\tilde{r},\theta, \varphi \right)=-\int \frac{\omega }{\sqrt{1-k \tilde{r}^2/a^2}} dt+R(\tilde{r})+P(\theta, \varphi).
\end{equation}

After we solve Maxwell's equations on the background of FRW Universe in Painlev\'{e} spacetime we solve for $\det\Xi = 0$, then we find the following radial part for the wave solution 
\begin{equation}
R(\tilde{r})_{\pm}=\frac{\omega}{\sqrt{1-k\tilde{r}^2/a^2}}\int dt - \int \frac{g(\tilde{r})E(\tilde{r})\pm\sqrt{g(\tilde{r})^2 E(\tilde{r})^2-f(\tilde{r})\Big[\Delta(\tilde{r}) \Big( (\partial_{\theta}P)^2+(\partial_{\varphi}P)^2\Big)-h(\tilde{r}) E(\tilde{r})^2\Big] }}{f(\tilde{r})}d\tilde{r}.
\end{equation}

By repeating the same argument as in the last section we end up with exactly the same expressions for the imaginary part
\begin{equation}
\text{Im}R_{+}=\pi \omega \tilde{r}_{A}, \,\,\,\,\,\text{Im}R_{-}=0.
\end{equation}

Using the Boltzmann formula $\Gamma=\exp\left(-\omega/T_{H}\right)$, we find the same result for the Hawking temperature
\begin{equation}
T_{H}=\frac{1}{2 \pi \tilde{r}_{A} }.
\end{equation}

Hence, we have concluded that the temperature of the Hawking radiation associated with the apparent horizon does not depend on the mass or type of tunnelling particles. It took Jacobson to show that Einstein's field equations can be interpreted as a thermodynamic law $TdS=dE+PdV$ \cite{jacobson}. Later on, this idea was put forward by many authors, in Refs. \cite{frw5,frw6} this idea was generalised to the Friedmann equation for FRW universe. Among other things, it was shown that Friedmann equation can also be viewed in terms of the first law of thermodynamics $dE=TdS+WdV$, at the apparent horizon.  Note that $W=(\rho-P)/2$, is the work density, $E$ is the energy, $V$ gives the volume inside the apparent horizon, $\rho$ is energy density, and $P$ is pressure of matter in the universe. Recently in Ref. \cite{frw7} a modified Friedmann equation with a power-law corrected entropy was investigated. We plan in the near future to investigate the role of entanglement of tunnelling particles from the apparent horizon.

\section{Conclusion}
In this article we have investigated the Hawking radiation of massive/massless vector particles in the framework of quantum tunnelling method from the apparent horizon of FRW universe. We have solved the relativistic Proca equation for a massive and massless Hawking quanta working in a Painlev\'{e}-type spacetime metric for FRW universe.  In the last part we have found the radial solution of the wave function using the WKB approximation method and the Kodama vector. We have shown that quantum tunnelling of massive/massless vector particles from the apparent horizon of FRW Universe give rise to the same expression for the Hawking temperature as in the case of scalar or Dirac particles. 

\section*{Competing Interests}
The authors declares that there is no conflict of interest regarding the publication of this paper.

\begin{acknowledgments}
This work was supported by the Chilean FONDECYT Grant No. 3170035 (A\"{O}).
\end{acknowledgments}


\begin{thebibliography}{0}



\bibitem{hawking1} S. W. Hawking, Commun. Math. Phys. \textbf{43}, 199
(1975); erratum-ibid, \textbf{46}, 206 (1976).


\bibitem{hawking2} Gibbons G W and Hawking S W 1977, Phys. Rev. D \textbf{15}%
, 2752

\bibitem{hawking3} G. W. Gibbons and S. W. Hawking, Phys. Rev. D \textbf{15}%
, 2738 (1977).


\bibitem{perkih1} M.K. Parikh, F. Wilczek, Phys. Rev. Lett. \textbf{85},
5042 (2000).

\bibitem{perkih2} M.K. Parikh, Phys. Lett. B \textbf{546}, 189 (2002).

\bibitem{perkih3} M.K. Parikh, Int. J. Mod. Phys. D \textbf{13}, 2351 (2004).

\bibitem{kraus1} P. Kraus, F. Wilczek, Mod. Phys. Lett. A \textbf{9}, 3713
(1994).

\bibitem{kraus2} P. Kraus, F. Wilczek, Nucl. Phys. B \textbf{437}, 231
(1995).

\bibitem{ang} M. Angheben, M. Nadalini, L. Vanzo, S. Zerbini, J. High Energy
Phys. \textbf{05}, 014 (2005).

\bibitem{sri1} K. Srinivasan, T. Padmanabhan, Phys. Rev. D \textbf{60},
024007 (1999).

\bibitem{sri2} S. Shankaranarayanan, K. Srinivasan, T. Padmanabhan, Mod.
Phys. Letts. \textbf{16}, 571 (2001)

\bibitem{vanzo} L. Vanzo, G. Acquaviva, R. Di Criscienzo, Class. Quantum
Gravity \textbf{28}, 18 (2011).

\bibitem{mann0} R. Kerner, R.B. Mann, Phys. Rev. D \textbf{73}, 104010
(2006).

\bibitem{mann1} R. Kerner and R.B. Mann, Class. Quant. Grav. \textbf{25},
095014 (2008).

\bibitem{mann2} R. Kerner and R.B. Mann, Phys. Lett. B \textbf{665}, 277-283
(2008).

\bibitem{mann3} Alexandre Yale, Robert B. Mann Phys. Lett. B \textbf{673},
168-172, (2009).



\bibitem{xiang1} Xiang-Qian Li, Ge-Rui Chen, Physics Letters B \textbf{751}
34-38, (2015)

\bibitem{kruglov1} S.I Kruglov, Mod. Phys. Lett. A \textbf{29}, 1450203
(2014).

\bibitem{sakalli1} I. Sakalli, A. Ovgun, Gen. Relativ. Gravit. \textbf{48}, 1 (2016)


\bibitem{sakalli2} I. Sakalli, A. Ovgun, Eur. Phys. J. Plus (2015) \textbf{130} 110

\bibitem{a1} A. Ovgun, Int J Theor Phys (2016) 55: 2919
\bibitem{a2} A. Ovgun, K. Jusufi, Eur. Phys. J. Plus (2016) 131: 177
\bibitem{a3} I. Sakalli, A. Ovgun, K. Jusufi, Astrophys Space Sci (2016) 361: 330

\bibitem{k1} Kimet Jusufi, Ali Ovgun, Astrophys Space Sci (2016) \textbf{361}, 207

\bibitem{k2} Kimet Jusufi, EPL, 116 (2016) 60013

\bibitem{k3} Kimet Jusufi, Ali Ovgun A. Int J Theor Phys (2017). \url{doi:10.1007/s10773-017-3317-7}

\bibitem{giddings} Steven B. Giddings, Phys. Lett.B, 754, 39 (2016)

\bibitem{xiang2} Xiang-Qian Li, Physics Letters B, 763 (2016) 80-86

\bibitem{frw1} Ran Li, Ji-Rong Ren, Dun-Fu Shi, 	Phys.Lett.B670:446-448,2009

\bibitem{frw2} Tao Zhu, Ji-Rong Ren, Ming-Fan Li, JCAP 0908:010,2009

\bibitem{frw3} Rong-Gen Cai, Li-Ming Cao, Ya-Peng Hu, Class.Quant.Grav.26:155018,2009

\bibitem{frw4} Tao Zhu, Ji-Rong Ren, Douglas Singleton, Int.J.Mod.Phys.D19:159-169,2010

\bibitem{jacobson} T. Jacobson, Phys. Rev. Lett. 75, 1260(1995 

\bibitem{frw5} M. Akbar, Rong-Gen Cai, Phys.Rev.D 75:084003, 2007

\bibitem{frw6} M. Akbar, Rong-Gen Cai, Phys.Lett.B 648:243-248, 2007

\bibitem{frw7} K. Karami, A. Abdolmaleki, N. Sahraei, S. Ghaffari, JHEP 1108:150, 2011


\bibitem{npb} D. Chen et al., Nucl. Phys. B (2017), http://dx.doi.org/10.1016/j.nuclphysb.2017.02.020

\bibitem{samantha} B.R. Majhi, S. Samanta, Ann. Phys. 325, 2410 (2010)

\bibitem{Siahaan} Haryanto M. Siahaan, Eur. Phys. J. C (2016) 76:139

\end{thebibliography}
\end{document}